# Forecasting the COVID-19 vaccine uptake rate: An infodemiological study in the US

Xingzuo Zhou[a]* and Yiang Li[b]

[a]Department of Economics, University College London, London, United Kingdom

[b]Social Research Institute, University College London, London, United Kingdom

*Corresponding author: E-mail: xingzuo.zhou.19@ucl.ac.uk

**Abstract**

A year following the initial COVID-19 outbreak in China, many countries have approved emergency vaccines. Public-health practitioners and policymakers must understand the predicted populational willingness for vaccines and implement relevant stimulation measures. This study developed a framework for predicting vaccination uptake rate based on traditional clinical data—involving an autoregressive model with autoregressive integrated moving average (ARIMA)—and innovative web search queries—involving a linear regression with ordinary least squares/least absolute shrinkage and selection operator, and machine-learning with boost and random forest. For accuracy, we implemented a stacking regression for the clinical data and web search queries. The stacked regression of ARIMA (1,0,8) for clinical data and boost with support vector machine for web data formed the best model for forecasting vaccination speed in the US. The stacked regression provided a more accurate forecast. These results can help governments and policymakers predict vaccine demand and finance relevant programs.

Keywords: public health; infodemiology; machine-learning; vaccine; forecast

**Introduction**

Approximately one year following the initial COVID-19 outbreak in Wuhan, China, two vaccines were approved for emergency distribution in the US: Pfizer-BioNTech (BNT162b2) and Moderna (mRNA-1273).[1] Recent real-world and clinical trials showed these vaccines as safe and effective against symptomatic infection, hospitalization, intensive care unit admission, and death even during the rising prevalence of the Delta variant.[2,3] As there is currently no fully approved treatment against SARS-CoV-2 implemented on a large scale, the use of vaccines is currently the only global public-health strategy stopping infection spread and appearance of potentially dangerous virus mutations.[4,5] Therefore, public-health policymakers and practitioners must understand and predict the future progress of vaccination.

Although the US Centers for Disease Control (CDC) publishes daily statistics on vaccines administered across the US, there is usually a 1–3-week delay for veracious data;[6] thus, traditional forecasting methods on vaccination based on clinical data lack accuracy. This adverse impact on decision-making processes results in further deaths and expiration of vaccine validity. Specifically, nations with surplus vaccines due to insufficient forecast on the vaccine demand were difficult to donate or swap vaccines with other nations and faced the risks of vaccine waste. So, the expiry date of many vaccines that were due to expire in a short time was extended without concrete empirical evidence.[7] To improve the forecast, researchers developed a new method: infodemiology, defined as the science of abstracting health-related content generated by internet users to improve public health.[8] Although online user-generated information is updated quickly, it is unpredictable compared with traditional statistical methods, and machine-learning is proving reliable with high predictive power, enabling the assessment of the future vaccination progress.[9] It will provide more information to public health practitioners and policymakers to ensure more optimal distribution while facilitating the implementation of interventions based on infodemiological approaches to reduce vaccine hesitancy.[10] Optimized management of vaccine distribution may further assist governmental decision-making on potential allotment of predicted vaccine excess supplies, such as for developing nations, which are now facing vaccine shortages.[11] Furthermore, the practice of monitoring vaccinations using both clinical and non-clinical web data can result in more accurate predictions on future vaccination rates in real-time while better elucidating any infodemiological nuances germane to vaccination behaviors behind the statistics. The analysis using Google Trends search interest in anti-vaccine terms revealed a concerning trend, and the burden of COVID-19 did not dissuade much vaccine hesitancy.[12] Therefore, our study provides evidence facilitating the development of prompter and more accurate public-health policies.[13,14] Importantly, public-health interventions to increase vaccination rates can be improved through a more accurate understanding of vaccination speed; this is a critical area of consideration to stop infection spread and the appearance of potentially dangerous virus mutations.

Previously, the autoregressive model (AR) and autoregressive integrated moving average (ARIMA) were widely applied for infection prediction and analysis of associated vaccination uptake,[15-17] including confirmed cases of COVID-19.[18] During COVID-19 vaccination, it was applied to fully vaccinated people's clinical data to predict future global vaccination rate.[19] Importantly, ARX, a similar model, can also be applied to non-clinical web data, in addition to clinical data, which may contain measurement errors. Non-clinical web data are the infodemiological data using user-generated information from electronic media such as statistics on online public-health-related interactions by individuals, including Tweets, comments, and search queries[20,21] to inform public-health policymaking.[22,23] While previous studies[24,25] applied the practice of selecting search queries for web data based on expert judgments in Sycinska-Dziarnowska et al.,[26] heavy reliance on human judgment is costly and difficult to justify quantitatively. Alternatively, researchers proposed using real-time correlations between query term frequencies and clinical reports to automatically select terms for future prediction.[27] Still, dynamic changes in future health events may render historically predictive terms unreliable for later contemporary predictions due to concept drift in machine learning. Müller and Salathé[28] showed that the use of pre-COVID-19 social media data to predict vaccine sentiments would systematically misclassify data for prediction during the pandemic. Our study develops another approach extracting query keywords based on their frequency of mention in individual Tweets related to vaccinations.

Recently, infodemiological studies and machine-learning algorithms were also implemented with non-clinical data to better predict vaccinations. One study used a supervised machine-learning algorithm with multivariate ordinary least squares (OLS) regression to explore the predictive power of both non-clinical personal attitudes toward scientific information and the clinical experience of severe respiratory disease on the influenza vaccination rate; accordingly, the area under the receiver operating characteristic curve (AUC) for this mixed clinical and non-clinical prediction method was 85%.[29] Carrieri et al.[30] implemented the supervised random-forest machine-learning algorithm on area-level indicators of institutional and socioeconomic backgrounds to predict the vaccine hesitancy rate for Italian local authorities, thus helping public-health practitioners run targeted awareness campaigns. Their findings suggested that non-clinical features had the highest predictive powers in the random-forest algorithm, with an AUC of 0.836. Besides these algorithms, Gothai et al.[31] proposed supervised machine-learning via the Holt-Winter model to obtain a prediction that captured seasonal variations in vaccination across the year to improve accuracy. While web data had similar strong predictive powers when monitoring the Middle East respiratory syndrome outbreak in South Korea, Shin et al.[32] noted statistically significant lag correlation coefficients higher than 0.8 between non-clinical variables of Google query keywords and Tweets from Twitter. This emphasizes the importance of using pertinent and accurate non-clinical web variables in conjunction with clinical data. Building upon the use of a single model for clinical and non-clinical data, Santillana et al.'s[33] evolved method used ensemble methods of stacking regression that combine separate outcomes from each model of different statistical classifiers based on labeled Tweets to predict vaccinations; they achieved considerable testing accuracy of 85.71% in the 10-fold cross-validation. Besides using a clinical or non-clinical data method solely, Hansen et al.[14] proposed a mixture method using two methods' predictions to attain higher prediction accuracy.

However, no relevant studies have implemented statistical and machine-learning predictions using both clinical and infodemiological web data on national-level COVID-19 vaccinations. Thus, we focused on finding more accurate ways of predicting COVID-19 vaccination rates by innovatively using both clinical and web information. This is a method-comparison study where not only the compared methods varied, but also the dataset used with each method. We innovate through an infodemiological approach, that is, in the use of non-clinical web data in addition to clinical data so far used in this field of research. Specifically, this study proposes to forecast COVID-19 vaccination rates using the AR/ARIMA model with clinical data and the OLS/least absolute shrinkage and selection operator (LASSO)/machine-learning methods with web data. Then stacked regressions are used to combine both to generate new predictions. This study uses the root-mean-squared error (RMSE) across all statistical and machine-learning methods to determine the best model. Eventually, more accurate predictions are anticipated to better explain vaccination behaviors.

**Materials and Methods**

*Data sources*

We obtained study data from January 2 to July 27, 2021. The study data, collected from CDC, is the daily first-dose vaccination in the United States. Regarding the outcome variable, Hansen et al.[14] proposed that the vaccination-to-expectation ratio is a more accurate measurement than the simple daily vaccination rate. It is defined as $\frac{daily\ first-dose\ vaccination}{number\ of\ people\ expected\ to\ be\ vaccinated\ in\ hundreds}$, where people expected to be vaccinated are all individuals in the US who have not received relevant vaccinations.

We considered daily records of the first COVID vaccine dose published by the CDC as clinical data, while the relative interest as searched using selected words from Google Trends was considered as web data.

We initially downloaded the "Covid Vaccine Tweets" dataset from Kaggle, which consisted of the texts of all the tweets related to COVID-19 vaccines. We then converted the Tweet texts into a corpus file to select words for the web data, which was preprocessed to remove irrelevant numbers, punctuations, symbols, and stop words. We extracted 68,409 features mentioned more than once and created a table of features ranked by their frequency of occurrence. Next, we qualitatively assessed the top 1,000 most frequent words and checked such queries' availability on Google Trends to select 12 words for each category of attitude (positive, negative, and neutral). To ensure that the added words were relevant to vaccination, we used the relative search volume of queries with the word "vaccine" before each identified keyword. Finally, to convert web searches into quantitatively analyzable "web data," we searched the relative interest in Google Trends from December 21, 2020, to July 27, 2021, in the US. As Google Trends allows up to five words per search, the reference word "Joker" was used to standardize the index of relative search volume across each search. We then added all standardized indices into three categories (Table 1).

The outcome variable is represented as *Daily_0*, and the altitude variables (positive, neutral, and negative) are represented as *pt_0, nt_0, ng_0* (Table 2).

*Overview of forecasting methods*

Previously, statistical models with clinical data are widely used to predict vaccine uptake rate. The performance of forecasts is hypothesized to be improved when they are incorporated into infodemiological datasets. Herein, we generated our predictions through three stages. The first stage is repeating the most common statistical model AR/ARIMA with clinical data. The second stage is implementing machine-learning approaches on infodemiological datasets where OLS is the reference of the machine learning algorithms. The third stage is to "stack" different combinations of predictions from the first two stages, similar to the forecasting averaging method in traditional statistics.

Table 1. Web data (search categories and related words)

| Attitude | Related Search Words | |
|---|---|---|
| Negative [ng_0] | vaccine fever | vaccine variant |
| | vaccine pain | vaccine restriction |
| | vaccine headache | vaccine reaction |
| | vaccine side effect | vaccine adverse |
| | vaccine death | vaccine risk |
| | vaccine cost | vaccine blood clot |
| Neutral [nt_0] | vaccine update | Moderna |
| | vaccine safety | vaccine used |
| | vaccine rate | vaccine information |
| | vaccine last | vaccine impact |
| | current vaccination | vaccine feeling |
| | Pfizer | vaccine effectiveness |
| Positive [pt_0] | vaccine available | vaccine cdc |
| | vaccine near me | vaccinate child |
| | vaccine registration | vaccine doses |
| | vaccine appointment | second dose |
| | vaccine booking | first dose |
| | vaccine location | vaccinated |

Table 2. Summary of raw data

| Variable | Obs | Mean | Std.Dev. | Min | Max |
| --- | --- | --- | --- | --- | --- |
| Day | 219 | 22,379 | 63.36403 | 22,270 | 22,488 |
| daily_0 | 219 | 1.501081 | 1.036814 | .0034924 | 4.534625 |
| ng_0 | 219 | 1.398338 | .5264653 | .4320988 | 3.76 |
| nt_0 | 219 | 9.035692 | 4.044199 | 2.837719 | 22.61224 |
| pt_0 | 219 | 8.836565 | 5.366234 | 1.356429 | 55.42065 |

*Forecast with clinical data*

AR/ARIMA models are widely applied to forecast future vaccine uptake rates solely using clinical data, and the reliability of applying AR/ARIMA on clinical data had been proven in history. This requires stationarity of time-series data; a given sequence is stationary if the joint probability distribution remains constant over time. The Dickey-Fuller (dfuller) and unit-root tests can be applied to confirm stationarity, and the dfuller test's null hypothesis is the existence of a unit root. If the null hypothesis is rejected, stationarity is satisfied.

As Table 3 shows, stationarity holds so that AR/ARIMA can be implemented without further differencing. Autocorrelation and partial correlation are necessary to identify our AR/ARIMA model. To estimate AR(p), we needed to choose the number of AR lags (p) as follows:

$$\hat{E}_c(t) = \mu + \sum_{i=1}^{p} \beta_i E(t-i). \qquad (1)$$

In clinical data analysis, $\hat{E}_c(t)$ is the expected value of daily vaccination-to-expectation ratio on day *t*, $\mu$ is the intercept, $E(t-i)$ is the value of daily vaccination-to-expectation ratio on day (*t-i*), and $\beta_i$ is the weight of the *i*th lag term. According to the principle of parsimony, we chose lags 1, 3, 5, 6, and 7 due to their positive partial autocorrelation out of the 95% confidence interval (Nau, 2020; Figure 1). The ARIMA model is required to further ensure accurate predictions. A non-seasonal ARIMA model is presented as ARIMA(p,d,q), where p represents AR lags, d represents differencing (I) lags, and q represents moving average (MA) lags. The first two symbols are identified as ARIMA(p,0,q), as stationarity holds without differencing, and AR(p) is identified earlier. Autocorrelation is required to identify MA lags.

Table 3. Dickey-Fuller test

| Variable | Degree of Integration | Test Statistic |
|---|---|---|
| daily_0 | I(0) | -4.541*** |

* $p < .10$ ** $p < .05$ *** $p < .01$

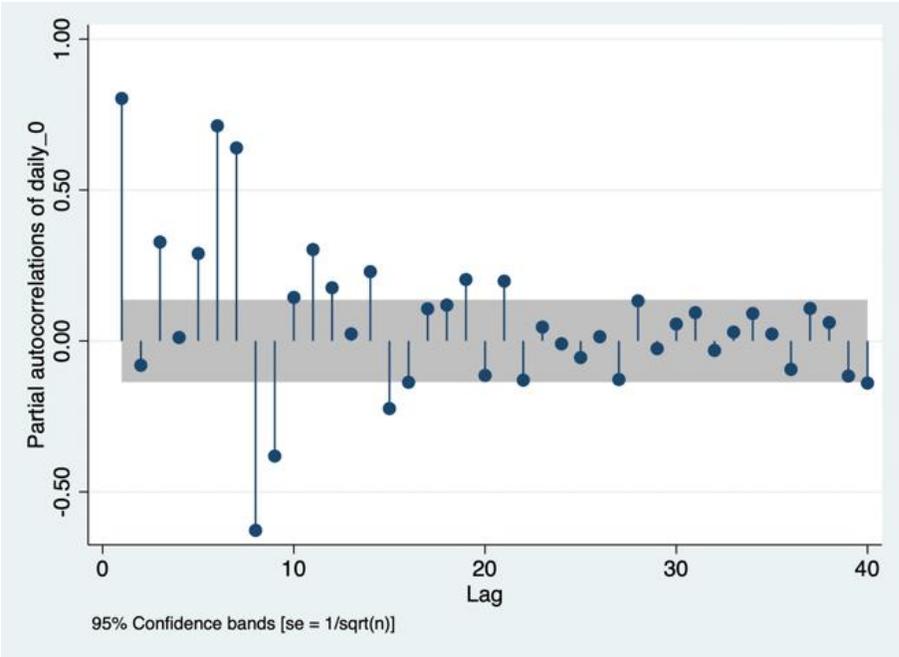

Figure 1. Partial autocorrelation plots.

According to the principle of parsimony, we chose lags 8, 15, 21, and 28 due to their autocorrelation out of the 95% confidence interval (Figure 2).[34] Therefore, ARIMA (p,0,q) estimates

$$\hat{E}_c(t) = \mu + \sum_{i=1}^{p} \beta_i E(t-i) + \sum_{k=1}^{q} \lambda_k e_{t-k}, \qquad (2)$$

where $e_{t-k}$ is the forecast error on day (*t-k*) and $\lambda_k$ is the weight of the forecast error. Further, ARIMA(p,0,q) uses q MA lags to smooth the data. Hence, ARIMA(1,0,8), ARIMA(1,0,15), ARIMA(1,0,21), ARIMA(1,0,28), ARIMA(3,0,8), ARIMA(3,0,15), ARIMA(3,0,21), ARIMA(3,0,28), ARIMA(5,0,8), ARIMA(5,0,15), ARIMA(5,0,21), and ARIMA(5,0,28) have been identified. The Akaike information criterion (AIC) and Bayesian information criterion (BIC) can be applied to determine the best-fit parameters of p and q.

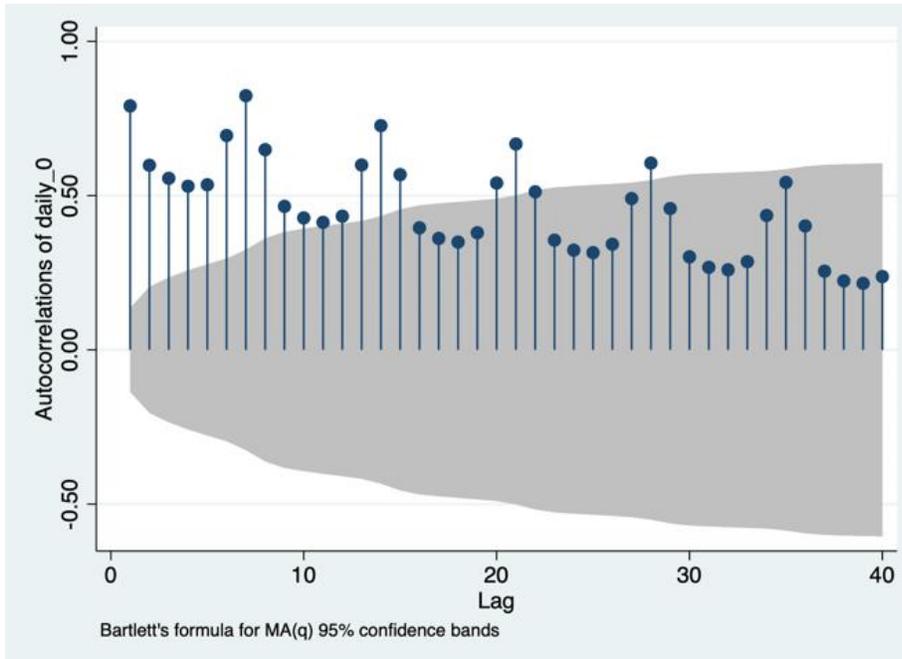

Figure 2. Autocorrelation plots. MA(q): moving average model of order q.

As Table 4 shows, ARIMA(7,0,28) had the smallest AIC, while ARIMA(7,0,8) had the smallest BIC and second smallest AIC. As the AIC of ARIMA(7,0,8) showed a very small difference compared with ARIMA(7,0,28) (according to the principle of parsimony), ARIMA(7,0,8) was identified as the best fit to our ARIMA model. Therefore, ARIMA(7,0,8) estimates

$$\hat{E}_c(t) = \mu + \sum_{i=1}^{7} \beta_i E(t-i) + \sum_{k=1}^{8} \lambda_k e_{t-k}. \qquad (3)$$

Briefly, the forecast for vaccination-to-expectation ratio on day *t* involves the weighted sum of vaccination-to-expectation ratios for the last seven days and the weighted sum of the forecast errors for the last eight days.

*Forecast with web data*

We applied linear regressions with OLS/LASSO, and machine-learning with boost and the random-forest algorithm to analyze comparably unpredictable web data. Machine-learning methods were implemented due to irregularity (e.g. outliers) of search frequencies.

*Linear regression with OLS*

Linear regression with OLS simply estimates

$$\hat{E}_w(t) = \alpha + \sum_{i=1}^{6} \beta_i x_i. \qquad (4)$$

In non-clinical web data analysis, $x_i$ represents the altitude, and its weight is given by $\beta_i$. Our independent variables were three altitudes and their one-period lags in the web data, so there were six independent variables. The one-period lag was decided because of a strong partial correlation between the two periods.

Table 4. Results of Akaike information criterion (AIC) and Bayesian information criterion (BIC)

| Models | AIC | BIC | Degrees of Freedom |
| --- | --- | --- | --- |
| ARIMA(1,0,8) | 208.5404 | 245.4628 | 11 |
| ARIMA(1,0,15) | 128.1185 | 188.537 | 18 |
| ARIMA(1,0,21) | 105.6378 | 182.8393 | 23 |
| ARIMA(1,0,28) | 72.20644 | 166.1909 | 28 |
| ARIMA(3,0,8) | 193.7227 | 234.0017 | 12 |
| ARIMA(3,0,15) | 132.6614 | 193.0799 | 18 |
| ARIMA(3,0,21) | 59.20707 | 139.7651 | 24 |
| ARIMA(3,0,28) | 39.18046 | 139.878 | 30 |
| ARIMA(5,0,8) | .437416 | 50.78621 | 15 |
| ARIMA(5,0,15) | -.5987349 | 73.24616 | 22 |
| ARIMA(5,0,21) | -1.744196 | 85.52705 | 26 |
| ARIMA(5,0,28) | 37.42017 | 141.4743 | 31 |
| ARIMA(6,0,8) | 2.424118 | 56.1295 | 16 |
| ARIMA(6,0,15) | 1.519492 | 78.72098 | 23 |
| ARIMA(6,0,21) | 3.681404 | 101.0224 | 29 |
| ARIMA(6,0,28) | -6.415284 | 114.4218 | 36 |
| ARIMA(7,0,8) | <u>-23.78244</u> | <u>33.27953</u> | 17 |
| ARIMA(7,0,15) | -17.77218 | 62.78589 | 24 |
| ARIMA(7,0,21) | -18.72024 | 81.97735 | 30 |
| ARIMA(7,0,28) | <u>-23.18934</u> | 94.29118 | 35 |

*Note*: ARIMA: autoregressive integrated moving average.

*LASSO regression*

The LASSO regression is very similar to linear regression but is more accurate for predictions. It aims to fit the best-fitting model of least bias by minimising the squared errors whilst avoids overfitting irrelevant features by selecting a reduced set of known covariates and reducing the size of coefficients. In this study, we used the same variables as in "*Linear regression with OLS*" to make predictions via LASSO.

*Boost classification*

Machine-learning was used to further ensure accurate prediction. In this process, we implemented the STATA module *r_ml_STATA*.[35] For all classifications, we used 10-fold cross-validation as the rule of thumb. We also used the first 212 days as training data and the last 7 days as testing data for all methods. The boost classification is an ensemble method that reduces errors by introducing a strong classifier from several weak classifiers, which is done by building an initial model for training data and then building models to minimize variance. A single robust model is based on many smaller models. The final prediction is a weighted sum of sequenced models known as weak classifiers.

*Random-forest classification*

Similarly, random-forest classification is also a machine-learning algorithm. It is another ensemble method referred to as the bagging algorithm together with featured randomness. It reduces the variance of individual classification trees by randomly selecting from the dataset. Averaging these uncorrelated predictors produces a final prediction using this algorithm.

**Forecast with mixed clinical and non-clinical web data model**

To combine clinical and non-clinical web data models, we use stacking regression. Specifically, we regress the actual vaccination-to-expectation ratio on predictions from the AR/ARIMA clinical model and predictions from one of four non-clinical web data models. Since there are two clinical models and four non-clinical web data models, eight combinations are available for each stacking method.

*Stacking with linear regression*

Linear regression is first used to "stack" predictions from the clinical data analysis $\widehat{E_c}(t)$ and web data analysis $\widehat{E_w}(t)$. The model estimates

$$\hat{E}(t) = \mu + \beta_c \widehat{E_c}(t) + \beta_w \widehat{E_w}(t). \tag{5}$$

$\hat{E}(t)$ is the expected value of daily vaccination-to-expectation ratio in day $t$, $\mu$ is the constant term, and $\beta_c$ and $\beta_w$ are weights of predictions based on clinical and non-clinical web data, respectively. Then, we used OLS to minimize the residual:

$$\sum_{t}^{n}(E(t) - \mu - \beta_c \widehat{E_c}(t) - \beta_w \widehat{E_w}(t))^2, \tag{6}$$

where $E(t)$ is the actual value of daily vaccination-to-expectation ratio on day $t$. We minimize differences between the actual and predicted daily vaccination-to-expectation ratios on day $t$.

*Stacking with support vector regression (SVR)*

We continued to "stack" predictions from clinical data analysis $\widehat{E_c}(t)$ and web data analysis $\widehat{E_w}(t)$, but with a different regression: SVR. SVR classification is a supervised machine-learning model that splits the dataset into two categories. We applied it to the continuous variable here using SVR, which generates a regression similar to the linear regression model. In contrast with the OLS method, we find the coefficients of SVR by minimizing the coefficient vector's norm:

$$\min_{\mu,\beta_c,\beta_w} \sum_{t}^{n} V\left(E(t) - \mu - \beta_c\widehat{E_c}(t) - \beta_w\widehat{E_w}(t)\right) + \frac{\lambda}{2}(\mu^2 + \beta_c^2 + \beta_w^2), \tag{7}$$

where $\lambda$ controls the penalty for large weights of clinical and non-clinical web data predictions. We defined $V(r) = |r| - \epsilon$, where $E(t) - \mu - \beta_c\widehat{E_c}(t) - \beta_w\widehat{E_w}(t)$ is represented by $r$, and $\epsilon$ is a hyperparameter that controls the maximum error of predictions allowed. Both $\epsilon$ and $\lambda$ allow us to define the tolerance level of error in our model. Unless $|r| < \epsilon$, $V(r)=0$.

To find the model with the least prediction error, RMSE is used to compare among traditional clinical, innovative infodemiological, and stacked models by calculating differences between predictions $\hat{y}_i$ and actual values $y_i$:

$$RMSE = \sqrt{\frac{(\hat{y}_i - y_i)^2}{N}}. \qquad (8)$$

A model with smaller RMSE has less prediction error and higher accuracy.

**Results**

As Table 5 shows, both clinical models have smaller RMSE than non-clinical models. Besides, ARIMA(7,0,8) had a smaller RMSE than AR(7). So, ARIMA(7,0,8) performed better than AR(7), and models with only clinical data performed better than models with only web data. Further, linear regression with OLS performed best across all web data methods.

Table 5. Root-mean-square error of forecast models based on single clinical or web data

|              | Clinical Data | Web Data |
|--------------|---------------|----------|
| AR(7)        | .3480         |          |
| ARIMA(7,0,8) | **<u>.2524</u>** |       |
| OLS          |               | **.9069** |
| LASSO        |               | .9253    |
| boost        |               | 1.024    |
| randomforest |               | .9485    |

*Note*: **Boldfaced** numbers: better than other models with same data; <u>underlined</u> numbers: better than all other models; AR: autoregressive; ARIMA: autoregressive integrated moving average; OLS: ordinary east squares; LASSO: least absolute shrinkage and selection operator.

As Table 6 shows, under stacking with both OLS and SVR, the combination of ARIMA(7,0,8) and boost performed the best, and stacking with SVR outperformed stacking with OLS. Further, the regression using SVR to stack ARIMA(7,0,8) and boost together has smaller RMSE than the traditional clinical data model ARIMA(7,0,8). So, stacked regression improves prediction accuracy.

Table 6. Root-mean-square error of forecast models based on ensemble models

|  | OLS | SVR |
| --- | --- | --- |
| AR + OLS | .3195 | .3506 |
| ARIMA + OLS | .2441 | .3019 |
| AR + LASSO | .3180 | .3322 |
| ARIMA + LASSO | .2438 | .2369 |
| AR + Boost | .3292 | .3478 |
| ARIMA + Boost | **.2405** | **.2294** |
| AR + Randomforest | .3197 | .2929 |
| ARIMA + Randomforest | .2425 | .3387 |

*Note*: **Boldfaced** numbers: better than other models with same data; underlined numbers: better than all other models; AR: autoregressive; ARIMA: autoregressive integrated moving average; OLS: ordinary east squares; SVR: support vector regression; LASSO: least absolute shrinkage and selection operator.

Table 7 shows 7-day predictions of the vaccination-to-expectation ratio. During the forecasting process, we removed actual vaccination-to-expectation ratios from July 21 to 27, 2021, so that this 1-week forecast was the "out-sample."

Table 7. Seven-day forecast of the best model compared with the actual data and traditional clinical model.

| Day | daily_0 (Actual) | ARIMA (Clinical) | SVR [ARIMA + Boost] |
|---|---|---|---|
| 2021-07-21 | 3.4929 | 3.2058 | 3.1999 |
| 2021-07-22 | 3.8026 | 3.2589 | 3.2807 |
| 2021-07-23 | 4.5346 | 3.5598 | 3.4643 |
| 2021-07-24 | 2.9873 | 2.4667 | 2.5136 |
| 2021-07-25 | 1.7189 | 1.5900 | 1.6573 |
| 2021-07-26 | 3.9796 | 2.9922 | 2.9534 |
| 2021-07-27 | 2.5781 | 3.4296 | 3.4175 |

*Note*: Best model: SVR [ARIMA + boost]; actual model: daily_0; traditional clinical model: ARIMA. SVR: support vector regression: ARIMA: autoregressive integrated moving average.

Figure 3 shows the forecast for the last 40 days. Prior to July 21, 2021, predictions were made based on all available data; beginning with July 21, 2021, actual data of the daily vaccination-to-expectation ratio were removed when forecasting.

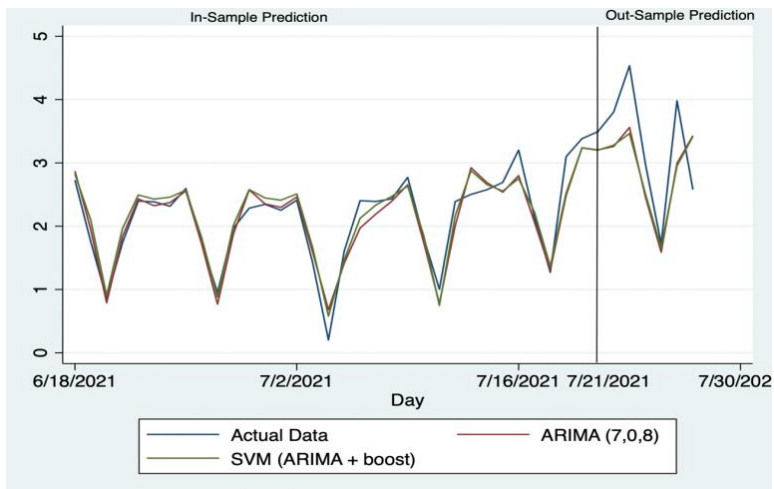

Figure 3. Actual data and predictions. SVR: support vector regression; ARIMA: autoregressive integrated moving average.

**Discussion**

Using web data, this study produced more accurate forecasts than traditional forecasting methods with solely clinical data, as reflected by RMSE, proving the function of online networks in predicting populational willingness to receive vaccinations. Specifically, compared with the traditional ARIMA model with clinical data, SVR with the predictions of both clinical data using ARIMA (7,0,8) and web data using Boost algorithm reduced RMSE by 9.1%. Figure 4 shows the vaccination uptake rate from December 21, 2020, to April 5, 2021. Over this period, the vaccination rate steadily increased, likely attributed to both the initial supply shortages and growing efficiency in vaccination facilities as rollout progressed. As Figure 5 shows, the vaccination-to-expectation ratio remained steady at approximately 2.2 from April 6, 2021, to June 30, 2021, likely attributed to two factors. First, on April 26, 2021, US President Joseph Biden announced that the US would begin donating vaccines to other countries.[36] This implied that the US was very likely to have sufficient vaccines for its whole population. Two, the number of new COVID-19 cases steadily declined over this period,[37] which alleviated populational concerns over the possibility of a more serious pandemic. As Figure 6 shows, the vaccination-to-expectation rate rose slowly and steadily from July 2 to 30, 2021. In this regard, the number of new COVID-19 infections rose sharply on July 1, 2021, which increased concerns that the pandemic would worsen. It might potentially be the factor leading populational willingness to receive vaccinations to increase. As Table 8 shows, these trends are also mathematically evident through Kendall's $\tau$. In sum, policymakers must understand daily vaccine demands in relation to vaccine supply and other factors when making relevant decisions.

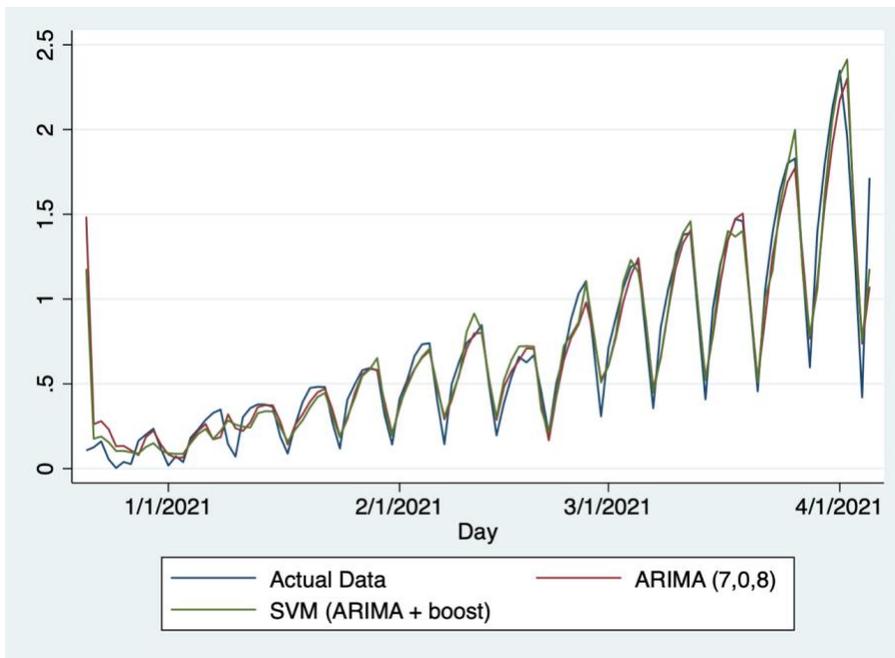

Figure 4. Actual data and predictions from December 21, 2020, to April 5, 2021. SVR: support vector regression; ARIMA: autoregressive integrated moving average.

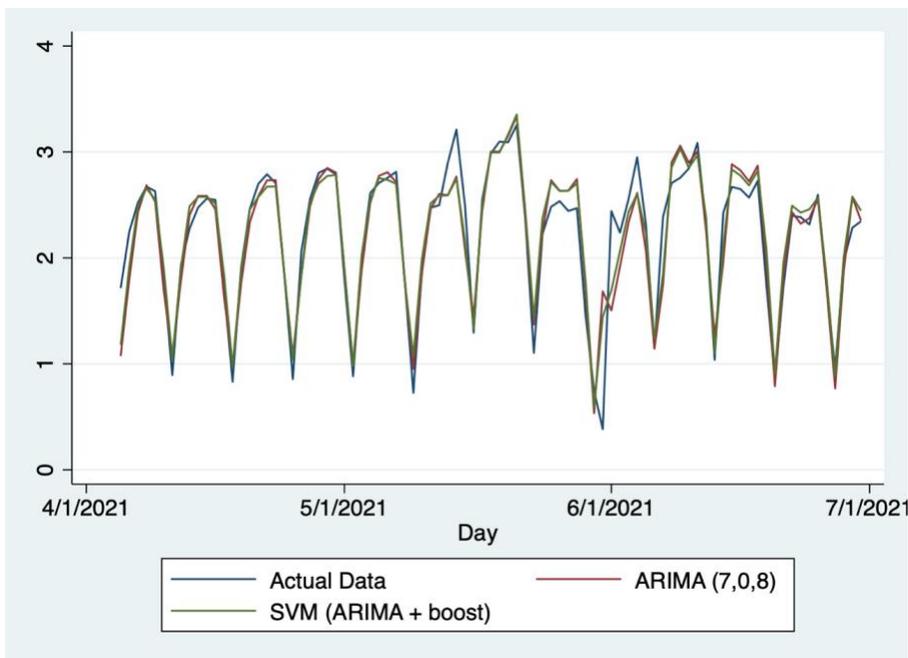

Figure 5. Actual data and predictions from April 6, 2021, to June 30, 2021. SVR: support vector regression; ARIMA: autoregressive integrated moving average.

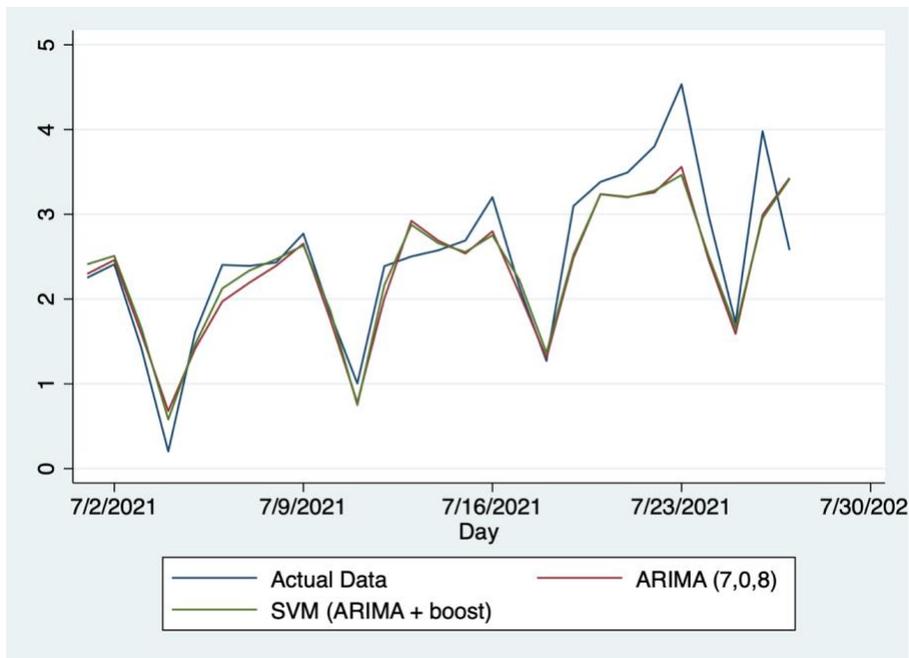

Figure 6. Actual data and predictions from July 1, 2020, to July 27, 2021. SVR: support vector regression; ARIMA: autoregressive integrated moving average.

Table 8. Kendall's $\tau$

| Period | Kendall's $\tau$ | Conclusion |
| --- | --- | --- |
| 2020-12-21–2021-04-05 | 0.6960 | Rising |
| 2021-04-06–2021-06-30 | -0.0326 | Steady |
| 2021-07-01–2021-07-27 | 0.4530 | Rising |

On July 27, 2021 (the last day considered in this study), it initially seemed that forecasts of the vaccination-to-expectation ratio were quite different from actual data in numbers and trends. Thus, forecasts appeared to overestimate this day and were higher than those from the previous day, whereas actual data showed a decline here. Despite this seemingly large error, forecasts may have provided some information based on potentially existing measurement errors by the CDC. In this regard, data gathered by the CDC are rarely of sufficient accuracy in the daily real-time context. Many such organizations and agencies globally implement the standard protocol of checking and adjusting data on later dates. The authors initially began this research in June but repeated the procedures multiple times until July 27, 2021. As the data provided by the CDC were adjusted several times over this period, it was observed that CDC also made measurement errors. Herein, the increased predictions from July 27, 2021 may signify that the CDC made positive measurement errors. This point may also help public-health practitioners and policymakers improve the accuracy of their policies.

Unfortunately, this research had a few unavoidable limitations. First, the US provides vaccinations to all individuals living in the country, regardless of citizenship status or permanent residency. Thus, temporary residents such as students also expected vaccinations. As these data were not available, adjustment based on the US population was made according to the Migration Policy Institute;[38] foreign nationals on temporary visas comprise around 7.1% of the total population, including citizens and permanent residents. Second, there was a slight error in the collected web data. As the relative interests of a few variables for a few days were too small to be illustrated, they were censored as "<1." To fix this, we simply adjusted them to 0.5, owing to the assumption of probability distribution. However, this adjustment should not have significantly impacted the forecasts, as the relative interests were very small and only affected a few days.

Despite unavoidable errors of no consequence, the forecasts in this study should help public-health practitioners and policymakers better foresee vaccine uptake behaviors and therefore develop more appropriate policies. Specifically, many sites such as Google currently collect personal information such as age and race when their users access the contents. Within the law, public-health practitioners and policymakers may predict vaccine uptake rate for different age, cultural and other groups more accurately with the help of those sensitive data. Further, relevant forecast models can also be applied to other countries and epidemic events in future settings. In addition, a new potential approach to encourage vaccination is appeared in this study: censoring negative social media contents. This study is based in US, so all contents are visible within the law. In the future, we aim to discover the impact of censoring negative social media contents through further studies in places where censoring negative social media contents exists.

Acknowledgements: We appreciate Dennis Kristensen, Dylan Kneale and editors from Taylor and Francis Editing for providing helpful suggestions.

Funding: No funding was used.

Conflict of interest: No conflict of interest is declared.

Data availability: Data that support the findings of this study are openly available in Covid Vaccine Tweets at https://www.kaggle.com/kaushiksuresh147/covidvaccine-tweets.

of Business, Duke University; 2020 Aug 18 [accessed 2021 Aug 25]. https://people.duke.edu/~rnau/411arim3.htm

35. Cerulli G. Machine Learning using Stata/Python. arXiv.org. 2021 Mar 3. [accessed 2021 Aug 25]: [22 p.]. arXiv: 2103.03122

36. Widakuswara P. Biden says US will donate 500 million COVID vaccines to world. Voice of America. 2021 Jun 10. [accessed 2021 Aug 25]. https://www.voanews.com/usa/biden-says-us-will-donate-500-million-covid-vaccines-world

37. Centers for Disease Control and Prevention, COVID Data Tracker. Trends in number of COVID-19 cases and deaths in the US reported to CDC, by state/territory Atlanta (GA): Centers for Disease Control and Prevention, US Department of Health and Human Services. 2021 [accessed 2021 Jul 28]. https://covid.cdc.gov/covid-data-tracker/#trends_dailytrendscases

38. Batalova J, Hanna M, Levesque C. Frequently requested statistics on immigrants and immigration in the United States. Washington (DC): Migration Policy Institute. 2021 Feb 11 [accessed 2021 Aug 25]. https://www.migrationpolicy.org/article/frequently-requested-statistics-immigrants-and-immigration-united-states-202029